\documentclass[twocolumn,aps,pra,superscriptaddress,nofootinbib,longbibliography]{revtex4-2}
\usepackage{graphicx}
\usepackage{dcolumn}
\usepackage{bm}
\usepackage[latin5]{inputenc}
\usepackage{amssymb,amsmath,amsthm,amsbsy,mathptmx}
\usepackage{amsfonts}
\usepackage{mathtools}
\usepackage{physics}
\DeclareMathAlphabet{\mathcal}{OMS}{cmsy}{m}{n}
\usepackage{tabularx}
\usepackage{xcolor}
\usepackage{mathtools}
\usepackage{oubraces}
\usepackage{dsfont}
\usepackage[title, titletoc]{appendix}
\usepackage{graphicx}
\usepackage{rotating}
\usepackage[most]{tcolorbox}
\usepackage{enumerate}
\usepackage{subfloat}
\usepackage[colorlinks=true,linkcolor=blue,citecolor=blue,urlcolor=blue]{hyperref}%
\usepackage{cleveref}
\usepackage[normalem]{ulem}
\Crefname{subfigures}{figure}{figures}
\Crefname{subfigures}{Figure}{Figures}

\begin{document}

\title{Entropy production in non-Markovian collision models: Information backflow vs system-environment correlations}

\author{H\"useyin T. \c{S}enya\c{s}a}
\affiliation{Department of Physics, Faculty of Science and Letters, Istanbul Technical University, 34469 Maslak, Istanbul, Turkey}

\author{\c{S}ahinde Kesgin}
\affiliation{TEBIP High Performers Program, Board of Higher Education of Turkey, Istanbul University, Fatih, \.{I}stanbul 34452, Turkey}

\author{G\"oktu\u{g} Karpat}
\email{goktugkarpat@ieu.edu.tr}
\affiliation{Faculty of Arts and Sciences, Department of Physics, \.{I}zmir University of Economics, \.{I}zmir, 35330, Turkey}

\author{Bar\i\c{s} \c{C}akmak}
\email{baris.cakmak@eng.bau.edu.tr}
\affiliation{College of Engineering and Natural Sciences, Bah\c{c}e\c{s}ehir University, Be\c{s}ikta\c{s}, \.{I}stanbul 34353, Turkey}



\begin{abstract}
We investigate the irreversible entropy production of a qubit in contact with an environment modelled by a microscopic collision model both in Markovian and non-Markovian regimes. Our main goal is to contribute to the discussions on the relationship between non-Markovian dynamics and negative entropy production rates. We employ two different types of collision models that do or do not keep the correlations established between the system and the incoming environmental particle, while both of them pertain to their non-Markovian nature through information backflow from the environment to the system. We observe that as the former model, where the correlations between the system and environment is preserved, gives rise to negative entropy production rates in the transient dynamics, the latter one always maintains positive rates, even though the convergence to the steady-state value is slower as compared to the corresponding Markovian dynamics. Our results suggest that the mechanism underpinning the negative entropy production rates is not solely non-Markovianity through information backflow, but rather the contribution to it through established system-environment correlations.

\end{abstract}

\maketitle


\section{Introduction}\label{intro}

The quest to describe and understand the dynamics of a quantum system that is interacting with its environment is a very important and central topic in physics that goes under the name of the theory of open quantum systems~\cite{BreuerPetruccione}. The mathematical structure of the traditional approach in the description of open quantum systems is in general quite involved both at the level of deriving the equations governing the system dynamics and at the level of solving them. Collision models, introduced as far back as in 1963~\cite{Rau}, provide an alternative approach that is simple, clear, and versatile to track the time evolution of the open system as well as, to some extent, the environment. In recent years, collision models are somehow ``re-invented"~\cite{ZimanDilutingQuantumInformation, ScaraniThermalizingQuantumMachines} and heavily utilized in addressing many different problems, ranging from modelling memory effects in open system dynamics~\cite{McCloskeyNonMarkovianitySystemenvironment2014,Ciccarello2013,Kretschmer2016,Cakmak2017,Campbell2018} to studying thermodynamics of quantum processes~\cite{StrasbergPRX,GabrieleNJP,FranklinPRL,GiacomoPLA}. We direct the interested reader to the following reviews and references therein~\cite{Ciccarello2021,Ciccarello2017,Campbell2021}.

In particular, we are interested in the relationship between non-Markovianity and the irreversible entropy production during open system dynamics, which has been a topic of intense debate over the last years~\cite{Kutvoven2015,PopovicEntropyProduction2018,Bhattacharya2017,Marcantoni2017,Argentieri2014,Strasberg2019,Xu2018,Gherardini2020,Bonanca2021,Ghosal2022,Li2021} and naturally extends to the discussions on the violations of the Landauer bound~\cite{Pezzutto2016,Lorenzo2015,LoFranco2019,Zhang2021,Bylicka2016}. It is a well-established fact that Markovian quantum dynamics, which can be described by a completely positive trace preserving (CPTP) map, lead to a positive rate of entropy production as proven by Spohn's inequality~\cite{Spohn}. On the other hand, in case of non-Markovian dynamics, the positivity of the entropy production rate is not always guaranteed due to the fact that conditions required by the Spohn's inequality to hold may not apply. In fact, it has been shown in Ref.s~\cite{PopovicEntropyProduction2018,Bhattacharya2017,Marcantoni2017,Argentieri2014} that non-Markovianity can lead to negative entropy production rates, while Ref.s~\cite{Strasberg2019,Xu2018} attempt to put forward a more general approach in understanding the relationship between these two phenomena. However, a unified and clear understanding of the relationship between non-Markovian dynamics and entropy production rate is still missing~\cite{Strasberg2019,LandiIrreversibleEntropy2021}. 

In this work, we consider a single thermal system qubit interacting with a thermal environment embodied by a stream of qubits at a temperature different than that of the system, in a collision model framework. By tuning the interactions between environmental qubits, we can control the non-Markovianity in the dynamics of the system qubit, where in the absence of such interactions we recover the Markovian limit. The intra-environment interactions result in non-Markovian dynamics through two mechanisms: \textit{i)} by passing the system information that flows into the environment particle to the parts of the environment which have not interacted with the system yet, but will do in the future, and \textit{ii)} in addition to \textit{(i)}, by establishing correlations between the system and aforementioned parts of the environment~\cite{tdbound}. Taking advantage of the versatility of the collision model framework, we employ two different strategies in the non-Markovian regime of the time evolution, in which we either keep the correlations established between the system and the next environmental unit or erase them. We examine the entropy production by describing the dynamics of our system using both of these strategies and try to pinpoint the cause of negativity in its rate by comparing the results we obtain.

The rest of the paper is organized as follows. In Sec~\ref{cm}, we describe the collision model framework and explain the two strategies using which we investigate the relationship between non-Markovianity and entropy production. We then continue with the introduction of the non-Markovianity measure we utilize and explicitly show that both of our strategies exhibit non-Markovian dynamics in Sec.~\ref{nonmarkov}. In Sec~\ref{ep}, we present our main results on the entropy production and its rate and discuss their behavior in relation to non-Markovian dynamics, considering the two different approaches we described. We point out several subtleties in calculating the entropy production in a collision model in Sec.~\ref{discuss} and conclude in Sec.~\ref{conclusion}.

\section{Collision Model}\label{cm}

Within the simplest, Markovian collision model framework, the dynamics of an open system particle $S$ is described by sequential and brief interactions (collisions) with an environment that consists of a stream of particles $\{E_i\}_{i=1}^N$. Following this interaction, the environmental particle is discarded and the system moves on to interact with a fresh environmental particle. Non-Markovianity can easily be introduced into this picture by adding an interaction between the environmental particles~\cite{McCloskeyNonMarkovianitySystemenvironment2014}, in particular between the particle that has just interacted with the system and the upcoming one. In this way, it becomes possible to partially recover the system information that is lost into the environment at a later time during the dynamics, which has been identified as one of the central mechanisms driving the memory effects as we will explain in the next section. The time evolution of the system can then be obtained iteratively repeating the steps described above and keeping track of the system state. The highly emphasized and appreciated versatility of collision models lies in this generality; one can change the number of particles in the system or environment, fix the type of the interactions between them in a suitable way for the problem under consideration, etc. 

Our model consists of a single system qubit and a stream of environment qubits all of which are initialised in a thermal state at difference temperatures with
\begin{align}
    \rho_0^S&=\frac{e^{-\beta_SH_S}}{\tr[e^{-\beta_SH_S}]}, & \rho_0^{E_i}=\frac{e^{-\beta_EH_E}}{\tr[e^{-\beta_EH_E}]},
\end{align}
where $H_S=\hbar\omega_S\sigma_z$, $H_E=\hbar\omega_E\sigma_z$ with $\sigma_z$ being the Pauli operator, and $\beta_S=1/k_BT_S$ and $\beta_E=1/k_BT_E$ are the inverse temperatures for the system and environment, respectively. The subscript of $\rho$ refers to the collision number, while its superscript is the particle label. Throughout this work, we will assume that the system and environment qubits are resonant, i.e. $\omega_S=\omega_E$. In addition, we also suppose that initially the global state of the system and the environment is in a product form given by $\rho_0^{SE}=\rho_0^S\bigotimes_{i=1}^N\rho_0^{E_i}$.

The time evolution is governed by the consecutive application of two unitary operators that describe the system-environment and the intra-environment interactions, where the presence of the latter is necessary for introducing non-Markovianity into the dynamics. In particular, both of these unitaries are given by a partial SWAP operator, which are expressed in the following form
\begin{align}\label{eq:interactions}
    \mathcal{U}_{S,E_i}(\nu) &= \cos(\nu) \mathds{1} + i \sin(\nu) \mathcal{S}, \\ \nonumber
    \mathcal{U}_{E_i,E_{i+1}}(\varepsilon) &= \cos(\varepsilon) \mathds{1} + i \sin(\varepsilon) \mathcal{S},
\end{align}
where $\mathds{1}$ is the $4\times4$ identity operator, and $\mathcal{S}$ is the SWAP operator in the energy eigenbasis ($\sigma_z$) of the two interacting particles given by
\begin{equation}
    \mathcal{S} =
    \begin{pmatrix}
        1 & 0 & 0 & 0 \\
        0 & 0 & 1 & 0 \\
        0 & 1 & 0 & 0 \\
        0 & 0 & 0 & 1 \\
    \end{pmatrix}.
\end{equation}
In Eq. (\ref{eq:interactions}), while the unitary evolution operator in the first line describes the system-environment interactions, the operator in the second line describes the intra-environment interactions, where $\nu$ and $\varepsilon$ characterize the strength of these interactions, respectively. When either $\nu$ or $\varepsilon$ attains the value $\pi/2$, one has a perfect SWAP operation between the two qubits. The intra-environment interaction strength $\varepsilon$ can be tuned to change the nature of the process from Markovian to non-Markovian, that is, memory effects can be introduced and controlled through this parameter. If no intra-environment collisions take place, i.e., $\varepsilon = 0$, then such a process corresponds to a Markovian one, also known as the quantum homogenization~\cite{ZimanDilutingQuantumInformation, ScaraniThermalizingQuantumMachines,McCloskeyNonMarkovianitySystemenvironment2014}.


Following Ref.~\cite{McCloskeyNonMarkovianitySystemenvironment2014,Campbell2018}, we consider two different strategies to describe the dynamics of our model in the non-Markovian regime. In both of the strategies we utilize in this work, on the $i^{\text{th}}$ step of the time evolution, the system-environment and intra-environment interactions take place by consecutive applications of $\mathcal{U}_{S,E_i}(\nu)$ and $\mathcal{U}_{E_i,E_{i+1}}(\varepsilon)$, respectively. The subtle, yet crucial, difference between the two strategies is whether the established correlations on the $(i-1)^{\text{th}}$ step between the system and $E_i$ is carried over to the $i^{\text{th}}$ step or not, which can be summarized as follows:
\begin{description}
    \item[Strategy 1] Before moving on to the next collision, we erase all the correlations between each ingredient of the model. In this case, the system state after the $i^{\text{th}}$ step of the considered model is given by
        \begin{equation}
        \rho_{i+1}^S=\tr_{E_i, E_{i+1}}\left[\mathcal{U}_{E_i,E_{i+1}}\mathcal{U}_{S,E_i}\left(\rho_i^S\otimes\tilde{\rho}_i^{E_i}\otimes\rho_i^{E_{i+1}}\right)\mathcal{U}_{S,E_i}^{\dagger}\mathcal{U}_{E_i,E_{i+1}}^{\dagger}\right],
    \end{equation}
    where $\tilde{\rho}_i^{E_i}=\tr_S[\rho_{i}^{SE_i}]$. Note that $\tilde{\rho}_i^{E_i}$ still contains some amount of information related to the system particle due to its interaction with the $(i-1)^{\text{th}}$ environment in the previous iteration.
    \item[Strategy 2] We keep the state of $\rho_{i}^{SE}$ untouched and use it as it is in the next iteration. Now, the system state after the $i^{\text{th}}$ step of the model is given as
    \begin{equation}
        \rho_{i+1}^S=\tr_{E_i, E_{i+1}}\left[\mathcal{U}_{E_i,E_{i+1}}\mathcal{U}_{S,E_i}\left(\rho_{i}^{SE}\otimes\rho_i^{E_{i+1}}\right)\mathcal{U}_{S,E_i}^{\dagger}\mathcal{U}_{E_i,E_{i+1}}^{\dagger}\right].
    \end{equation}
    Clearly, in such a case, any correlations established between $S$ and $E_i$ on the $(i-1)^{\text{th}}$ step, i.e. before they directly interact, is carried over to the $i^{\text{th}}$ step.
\end{description}
The mechanism behind the non-Markovianity has two different origins: \textit{(i)} the system information passed on to the incoming environment, and \textit{(ii)} correlations that are established between the system and the incoming environment. While in Strategy 1 the only mechanism for information backflow, i.e. non-Markovianity, is \textit{(i)}, in Strategy 2 we have contributions coming from both \textit{(i)} and \textit{(ii)}.

\section{Quantifying non-Markovianity}\label{nonmarkov}

Characterization and quantification of non-Markovianity in open quantum systems have been topics of very broad interest for more than a decade now \cite{nonmarkovRMP,Rivas2014}. Numerous measures of non-Markovianity have been proposed in the recent literature which claim to identify the memory effects in the dynamics of open systems using various different techniques.

In our study, we consider one of the most well-known and widely used non-Markovianity quantifiers known as the BLP measure~\cite{BreuerMeasureDegree2009}. In this approach, memory effects originating from the non-Markovian character of the open system dynamics is recognized through the distinguishability of open system states based on the trace distance between them. The trace distance between two density operators $\rho_1$ and $\rho_2$ is defined as 
\begin{equation}
D(\rho_1, \rho_2)\!=\!
\frac{1}{2}
||\rho_1\!-\!\rho_2||_1
\!=\!
\frac{1}{2}
\tr \left[(\rho_1\!-\!\rho_2)^{\dagger} (\rho_1\!-\!\rho_2)\right]^{1/2},
\label{trdist}
\end{equation}
where $||.||_1$ is the trace norm. In particular, the variations in the distinguishability between two arbitrary initial open system states throughout the dynamics are interpreted as the flow of information between the open system and its surrounding environment. If the distinguishability between two arbitrary initial states of the open system monotonically decreases during the dynamics of the system, i.e., $dD/dt<0$, then there is a one-way loss of information from the open system to the environment, which indicates a memoryless and thus Markovian evolution. On the other hand, if the distinguishability undergoes temporary revivals throughout the time evolution such that $dD/dt>0$, it means that there exists a backflow of information from the environment to the open system, giving rise to the non-Markovian memory effects in the dynamics.

\begin{figure}[t]
\centering
\includegraphics[width=1\columnwidth]{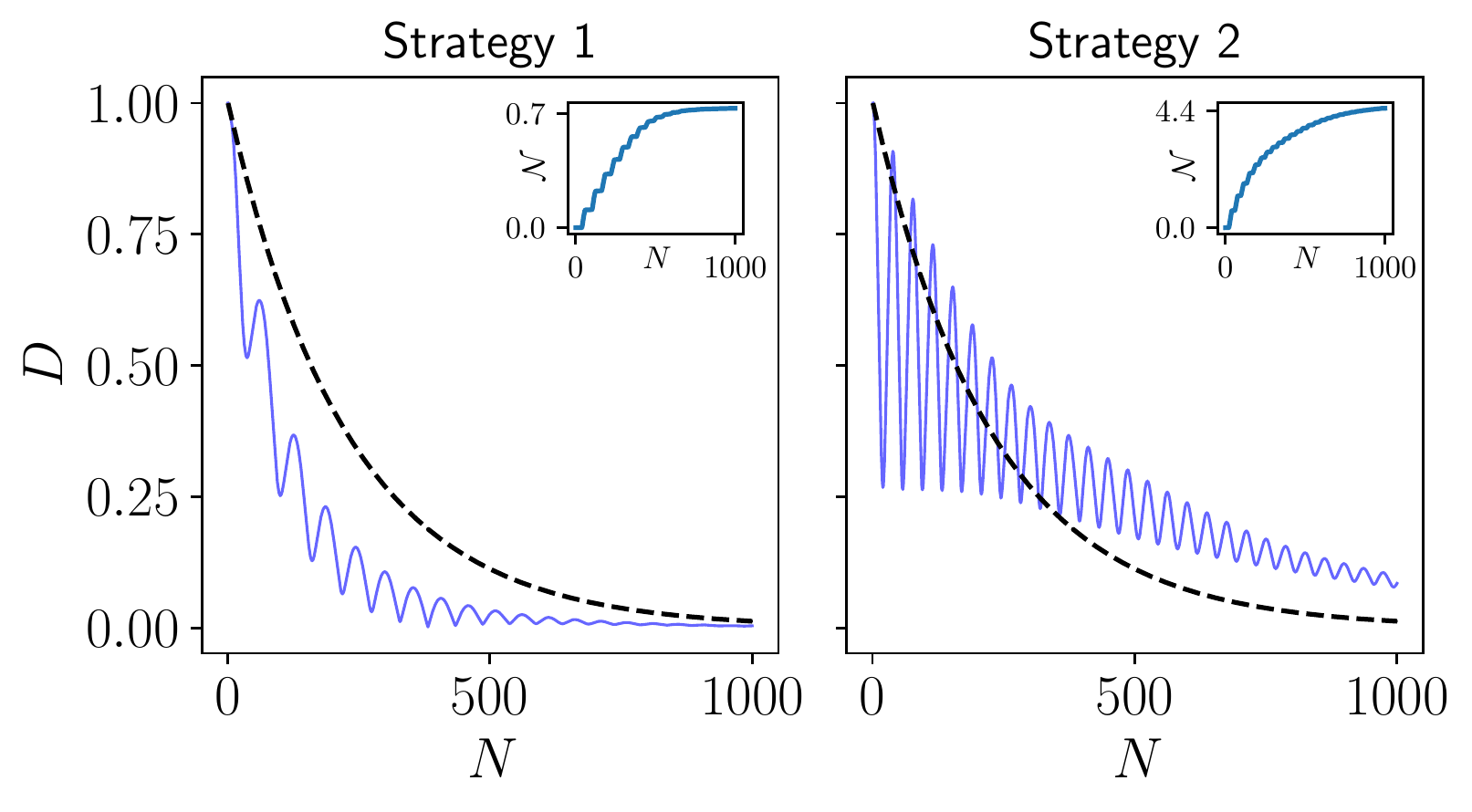}
\caption{Trace distance as a function of the number of collisions. While the dashed black lines show the behavior of the Markovian limit of both strategies, solid, blue lines display the behavior of the trace distance in the non-Markovian regime. Insets in both of the plots show the non-Markovianity measure, which is the sum of the amount of increases in the trace distance throughout the dynamics. The model parameters are chosen as $T_E=1$, system-environment interaction strength $\nu=0.05\times\pi/2$, and all particles in the model are resonant $\omega_S=\omega_E=1$. For non-Markovian dynamics intra-environment interaction strength is chosen as $\varepsilon=0.95\times\pi/2$. The initial state pair used in the calculation of the non-Markovianity measure is given by the eigenstates of the Pauli operator $\sigma_x$.}
\label{fig:tracedistance}
\end{figure}

On the basis of the above-mentioned interpretation, the degree of non-Markovianity of an open quantum system dynamics can be quantified as~\cite{BreuerMeasureDegree2009}
\begin{equation}
{\cal{N}}= \max_{\rho_1(0),\rho_2(0)} \int_{\dot{D}>0}\frac{dD}{dt}dt,
\label{NonM}
\end{equation}
where the optimization in the above equation should be performed over all possible initial state pairs $\rho_1(0)$ and $\rho_2(0)$ of the open system in principle. Since we will consider a collision model to describe the open system dynamics in our work, the dynamics occurs in discrete time steps. As a matter of fact, here we evaluate the degree of non-Markovianity using a discretized version of Eq.~(\ref{NonM}) as first considered in Ref.~\cite{BreuerDiscMeasure}
\begin{equation}
{\cal{N}}= \max_{\rho^s_{1,0},\rho^s_{2,0}} \sum_i \left[D(\rho^s_{1,i},\rho^s_{2,i})-D(\rho^s_{1,i-1},\rho^s_{2,i-1}) \right],
\label{NonMDisc}
\end{equation}
where the index $i$ denotes the collision number in the model.

In Fig. \ref{fig:tracedistance}, we display the behavior of the trace distance considering the two strategies we investigate in this work. Clearly, both of them exhibit non-Markovian behavior, as expected from earlier works~\cite{McCloskeyNonMarkovianitySystemenvironment2014,Cakmak2017}, with the degree of non-Markovianity being significantly higher in Strategy 2 due to the nature of the dynamics that keeps the established system-environment correlations. We are now ready to study the behaviour of the entropy production in both of these non-Markovian models that are different in a slight, but crucial way.

\section{Entropy Production}\label{ep}
As we have outlined in the description of our model, we have an initial thermal system state that is interacting with a thermal environment through sequential collisions. It is important to once again emphasize that, the system and the environment states are initially uncorrelated. The irreversible entropy production in accordance with the second law of thermodynamics can then be expressed as follows
\begin{equation}\label{eq:ep}
    \Sigma=\Delta S+\beta\Delta Q
\end{equation}
where $\Delta S=S(\rho_N^S)-S(\rho_0^S)$ with \(S(\rho) = -\tr(\rho \log{\rho})\) reflects the change in the von Neumann entropy, and $\Delta Q=\tr[H_E(\rho^E_{N}-\rho_0^{E})]$ is the heat exchanged with the environment. Here,  $\rho_0^S$ and $\rho_N^S$ respectively denote the initial state of the system and the state of it after the $N^{th}$ collision takes place. Specifically, this quantity characterizes the produced entropy in the system throughout its dynamics that cannot be traced back to reversible heat flow, and therefore is a measure of irreversibility of the dynamics. For the reasons we will address after presenting our results, below we first show that our model allows us to calculate the entropy production using an equivalent expression that only refers to the state of the system, i.e., without the necessity of performing any calculations involving the state of the environment particles.

The dynamical map considered in this work falls inside a very restricted class of maps which are known as thermal operations~\cite{Janzing2000,Brandao2015}. Such operations are characterized by the presence of a thermal environment and a global fixed point for the system state such that $U(\rho^S_*\otimes\rho^E)U^{\dagger}=\rho^S_*\otimes\rho^E$ is satisfied. Once the former criteria is met, it becomes possible to guarantee the existence of a global fixed point if the system-environment interactions are strictly energy preserving, i.e. $[U, H_S+H_E]=0$. Each iteration of our model, consisting of consecutive applications of two unitary operations, describe the system-environment and the environment-environment interactions. It is possible to show that the strict energy conservation criterion is satisfied at each step of the described dynamics, $[\mathcal{U}_{E_{n+1},E_n}(\nu)\mathcal{U}_{S,E_n}(\nu),H_S+H_E]=0$, since we also consider resonant system and environment qubits. Therefore, the global fixed point for our dynamical map is the initial thermal state of the environment, that is, $\rho_*^S=\rho_0^{E_i}$. In other words, independently of whether the dynamics of our system is Markovian or non-Markovian, the steady-state of the system ends up being the initial state of the environment qubits. Considering thermal operations, and also assuming that the global system-environment state is evolving unitarily, one can express the entropy production by referring only to the state of the system in the following way~\cite{LandiIrreversibleEntropy2021}
\begin{align}\label{eq:ep_main}
    \Sigma &=\sum_{i=1}^N\Sigma_i=\sum_{i=1}^N S(\rho_{i-1}^S || \rho_*^S)-S(\rho_i^S || \rho_*^S) \\ \nonumber
    &= S(\rho_0^S || \rho_0^{E_i})-S(\rho_N^S || \rho_0^{E_i})
\end{align}
Here, $S(\rho || \sigma)=\tr(\rho \log{\rho})-\tr(\rho \log{\sigma})$ is the quantum relative entropy. Due to the discrete nature of the evolution of our model, we expressed the total entropy production as a sum of the entropy produced at each step, $\Sigma_i$. The entropy production as given by Eq. (\ref{eq:ep_main}) will be the central quantity in this work. The above expression also holds when the system is weakly coupled to the environment~\cite{BreuerPetruccione}, which is, to a certain degree, similar to the strict energy conservation condition.

\begin{figure}[t]
\centering
\includegraphics[width=1\columnwidth]{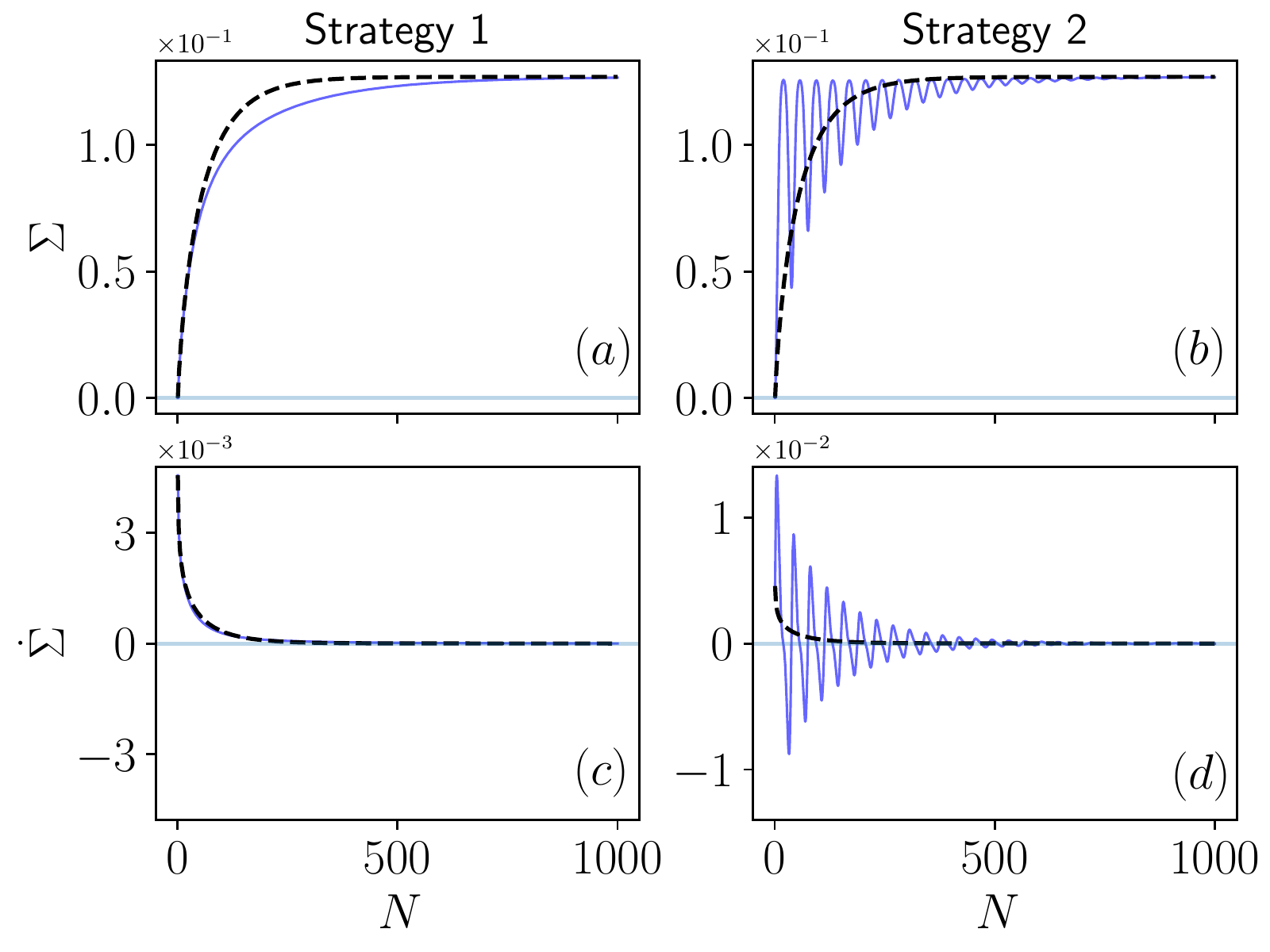}
\caption{Total irreversible entropy production {\bf (a)} and {\bf (b)}, and its rate {\bf (c)} and {\bf (d)} for Markovian (blacked dashed lines) and non-Markovian (solid blue lines) evolution. The model parameters are chosen as $T_S=0.1$, $T_E=1$, system-environment interaction strength $\nu=0.05\times\pi/2$, and all particles in the model are resonant $\omega_S=\omega_E=1$. For non-Markovian dynamics intra-environment interaction strength is chosen as $\varepsilon=0.95\times\pi/2$.}
\label{fig:main}
\end{figure}

An equally important quantity is the entropy production rate, $\dot{\Sigma}=d\Sigma/dt$. We follow a discretization approach for the time derivative of the entropy production similar to what we have done for the time derivative of the trace distance in the non-Markovianity measure calculation, and write it as
\begin{equation}
    \dot{\Sigma}=\Sigma_i-\Sigma_{i-1}.
\end{equation}
The entropy production rate $\dot{\Sigma}$ is always a positive quantity for Markovian dynamics due to the Spohn's inequality~\cite{Spohn}, which can also be seen from the fact that quantum relative entropy is a non-increasing function under completely positive trace preserving (CPTP) maps~\cite{Lindblad1975}. However, for non-Markovian dynamics it has been shown in a number of works that the entropy production rate can attain negative values throughout the transient dynamics~\cite{Argentieri2014,Kutvoven2015,Bhattacharya2017,Xu2018,Marcantoni2017,PopovicEntropyProduction2018,Strasberg2019,Li2021,Ghosal2022}. In fact, in the vast majority of these studies, the negativity of the entropy production rate $\dot{\Sigma}$ is claimed to be linked to the presence of memory effects due to non-Markovianity. In what follows, we will show that non-Markovianity originating from the information backflow alone is not sufficient to obtain negative entropy production rates, and the correlations between the system and the upcoming environmental particles are essential.

Fig.~\ref{fig:main} presents our results on the entropy production in both Markovian and non-Markovian regimes of the collision models described by Strategy 1 and Strategy 2. We begin our discussion with Strategy 1 in which, when the intra-environment collisions are introduced, non-Markovianity of the dynamics is only due to the backflow of system information that is transferred to the upcoming environment. In Fig.~\ref{fig:main} {\bf (a)} we display the total entropy production $\Sigma$ as given by Eq. (\ref{eq:ep_main}) both in Markovian (dashed line) and non-Markovian cases (solid line) which can observed to be a monotonic as a function of the collision number. This is clearly reflected in the rate of the entropy production $\dot{\Sigma}$ which remains positive throughout the dynamics, as shown in Fig.~\ref{fig:main} {\bf (c)} . Eventually, at the steady-state, the entropy production $\Sigma$ for the non-Markovian dynamics converges to that of the Markovian limit (dashed line), which is an expected result. The only difference between the Markovian and non-Markovian cases here is the slight slowing down of the entropy production resulting in the delay of the saturation at the steady-state in the latter one. 

We now turn our attention to the behavior of the same quantities in Strategy 2 displayed in Fig.\ref{fig:main} {\bf (b)} and {\bf (d)}. As expected, the Markovian limit of this case is identical to that of the Strategy 1, since the two strategies become identical when the intra-environment collisions are absent. However, there is a striking difference in the non-Markovian limit, where the entropy production $\Sigma$ shows fast oscillations in the transient time before flattening and saturating to the value at the steady-state. Obviously, the highly non-monotonic behavior of entropy production is reflected in its rate, causing $\dot{\Sigma}$ to attain negative values prior to converging to zero when the system is relaxed into the thermal state at the environment temperature.

We would like to elaborate on the reason why we have obtained different results between these two strategies. Despite the undoubtedly non-Markovian character of both versions of the collision model, the only difference between them is whether to keep the correlations between the system and the incoming environment or not. To support this point, after each step of the dynamics, we have checked the reduced states of the incoming environments in both strategies and seen that they are completely identical. This implies that, throughout the dynamics the local state of the environmental qubit just before its interaction with the system is identical in Strategy 1 and Strategy 2. The only difference between these two approaches is the fact that incoming environment is not correlated with the system in the former, and correlated with the system in the latter. Therefore, we conclude that the cause of negative entropy production rates cannot solely be the non-Markovian character of the dynamics. The crucial ingredient here is the correlations between the system and parts of the environment that has not interacted with it yet, but will do in the future. Note that, this is different from the system-environment correlations established \textit{after} the environment qubit has interacted with the system. 

It is interesting to note that in \cite{StevePrecursors} the authors bound the revivals of the trace distance from above at time $t$, using quantities involving established system-environment correlations at time $s$ with $t>s$, called precursors of non-Markovianity. They explicitly demonstrate the implications of their bound using a collision model that has the same features as our Strategy 2, i.e. when the system and incoming environment particle is correlated. As a result, the idea of precursors can also be useful in the context of detecting negative entropy production rates in non-Markovian dynamics.

Finally, we would like to comment on the particular choice of parameters that our calculations are based on. We deliberately consider a small system-environment interaction strength throughout this work. The reason behind this is to remain in the weak coupling regime, in order to avoid any peculiarities of thermodynamics in the strong coupling regime~\cite{Rivas2020}. The intra-environment interaction strength is set such that we are in the non-Markovian regime of both model, in order to be able to make a meaningful comparison that leads to the main conclusion of this work. In fact, for $\epsilon \leq 0.92 \times \pi/2$ Strategy 1 is becomes Markovian, therefore that region is not of interest for the purposes of this work. Similarly, when the initial environment temperature $T_E \geq 4.0$, Strategy 1 is no longer non-Markovian. Hence, within the parameter regions where both models admit a non-Markovian dynamics, only limitation of our results comes from the weak system-environment coupling assumption. Apart from that, although the behavior of the non-Markovianity and entropy production may show quantitative differences, their qualitative behavior remain the same for different choice of parameters, and our results remain intact.

\section{Discussion}\label{discuss}

In this section, we would like to highlight some subtle points in the calculation of the entropy production in a collision model framework. There are in fact equivalent but different mathematical expressions that the traditional expression for the entropy production given in Eq.~(\ref{eq:ep}) can be cast into. One of the most important expressions has been introduced in Ref.~\cite{EspositoEntropyProduction2010}, and further developed in~\cite{ReebWolf}, which allows us to express the entropy production in a purely information theoretic form as follows
\begin{equation}\label{eq:RW}
    \Sigma = \mathcal{I}_{\rho^{\prime}_{SE}}(S:E) + S(\rho^{\prime}_E || \rho_E),
\end{equation}
where \(\rho^{\prime}_{E} = \tr_S(\rho^{\prime}_{SE})\), $\mathcal{I}(\rho_{AB})=S(\rho_A)+S(\rho_B)-S(\rho_{AB})$ is the mutual information between bipartitions $A$ and $B$ of a quantum system. The above expression can be derived by assuming that the system and environment is initially in a product state, and their global evolution is described by a unitary operator. This form reflects two distinct contributions to the entropy production. First one is stemming from the correlations established between the system and the environmental particles throughout the dynamics, while the second one measures how much the environment is kicked out from equilibrium after interacting with the system. It is important to note that $\rho_{E}^{\prime}$ appearing in Eq.~(\ref{eq:RW}) refers to the whole environmental state. However, due to the nature of the collision model approach we follow in this work, each particle that has played its part in the description of the dynamics by interacting with the system (and with the next environment if intra-environment interaction is turned on) is discarded. As a result, at the end of the dynamics, we do not have access to the whole environment to calculate the mutual information that it shares with the system particle. 

On the other hand, note that the mutual information shared between the system and an environmental particle that has interacted with it asymptotically approaches to zero as the system particle continue further into the dynamics~\cite{Campbell2018}. This implies that the effect of discarding the environment particles that interacted with the system particle early on in the dynamics will have smaller contribution to the mutual information term. However, since discarding a quantum system never increases the mutual information~\cite{nielsen}, one can only get a lower bound on the entropy production from Eq.~(\ref{eq:RW}) with the present setting. Furthermore, the intra-environment collisions we have introduced to simulate a non-Markovian evolution complicates the calculation of the second term appearing in Eq.~(\ref{eq:RW}). The influence that changes the state of an environmental qubit is not just its interaction with the system but also the intra-environment interactions that have a non-negligible effect. One may attempt to circumvent this problem by keeping track of the change in the state of the both environmental particles that enter into the picture in a single iteration of the model. Nevertheless, it is important to note that this would only be a partial solution to the problem of discarding the environment particles.

It is also important to note that there are two critical assumptions behind the derivation of Eq.~(\ref{eq:RW}): initially factorized system-environment state and global unitary dynamics, which allows the usage of the invariance of the von Neumann entropy under unitary operations. In this work, we indeed start from a factorized initial system-environment state and the evolution is governed by unitary operators acting on the them. However, especially in Strategy 1, we interrupt the overall unitarity of the process by performing partial trace operations throughout the dynamics. In this sense, it is in general not possible to trace the state of the system+environment back to its initial state, which violates a very crucial assumption in the derivation of Eq.~(\ref{eq:RW}). 

Finally, we would like to mention that it is possible to reach the same conclusion that we did regarding the cause of the negative entropy production rates using the traditional form of the entropy production in Eq.~(\ref{eq:ep}), by addressing the second term from the state of the environment, but with caution. Clearly, the heat exchange (or the entropy flux) between the system and the environment only takes place when they are in contact. At this point, we have access to the state of the environmental qubit, which allows us to simply add up the contributions coming from the entropy flux at each step of the model. The crucial point here is the fact that one needs to look at the change in the state of both environment qubits that are involved in the description of the dynamics at that particular point, i.e. $\rho^{E_iE_{i+1}}$ at the $i^{\text{th}}$ step. Calculating the entropy flux from the change in the state of the environment qubit that is directly interacting with the system results in incorrect outcomes, which do not agree with the Markovian limit of the model.

\section{Conclusions}\label{conclusion}

We have investigated the relationship between the non-Markovianity of the dynamics and the behavior of the entropy production during the relaxation of an open system qubit to a thermal environment using a collision model approach. Taking advantage of the freedom provided by the collision model framework, we have considered two distinct strategies to describe the open system dynamics. In particular, we have either carried over or completely erased the correlations, which have been indirectly established between the system and the parts of the environment that have not yet directly interacted with the system. While the latter resulted in a monotonic behavior in the entropy production implying a positive rate, the former is highly non-monotonic giving rise to a negative entropy production rate in the transient time, before the system relaxes to the temperature of the environment. This points out to an important connection between non-Markovianity and entropy production rate: non-Markovian character of the open system dynamics alone is not sufficient to result in a negative entropy production rate, the presence of system-environment correlations is also necessary. Even though the dynamical model we consider in this work is limited to qubit system and environment particles, and thermal operations, we believe that our results has the potential to contribute to the discussions on the origin of the negative entropy production rates during non-Markovian open system dynamics, which is an open problem to date. 

\acknowledgements

B. \c{C}. and G. K. is supported by The Scientific and Technological Research Council of Turkey (TUBITAK) under Grant No. 121F246. B. \c{C}. is also supported by the BAGEP Award of the Science Academy. The authors would like to thank Steve Campbell, Gabriel T. Landi, and Mauro Paternostro for various discussions. 

\bibliography{biblio}

\end{document}